\documentclass[final,aps,prb,twocolumn,superscriptaddress,showpacs]{revtex4-1}
\usepackage{graphicx}% Include figure files
\usepackage{dcolumn}% Align table columns on decimal point
\usepackage{bm}% bold math
\usepackage{xr}
\usepackage{soul,xcolor}
\setstcolor{red} % set the color of cross-out line. This requires \usepackage{soul,xcolor}
\usepackage{kotex}
\begin{document}

\preprint{}

\title[]{TBA-enabled spin-coating of a percolatively connected GO nanosieve for thru-hole epitaxy: tuning GO flake stacking and coverage to control GaN nucleation}
% Force line breaks with \\
%\thanks{footnote to title of article}

\author{Gunhoon Beak}
\affiliation{Department of Physics and Research Institute for Basic Sciences, Kyung Hee University, 26 Kyungheedae-ro, Dongdaemun-gu, Seoul 02447, Republic of Korea}

\author{Changwook Dong}
\affiliation{Department of Information Display, Kyung Hee University, 26 Kyungheedae-ro, Dongdaemun-gu, Seoul 02447, Republic of Korea}

\author{Minah Choi}
\author{Jieun Yang}
\affiliation{Department of Chemistry and Research Institute for Basic Sciences, Kyung Hee University, Seoul, 02447, Republic of Korea}

\author{Joonwon Lim}
\affiliation{Department of Information Display, Kyung Hee University, 26 Kyungheedae-ro, Dongdaemun-gu, Seoul 02447, Republic of Korea}

\author{Chinkyo Kim}
\email{ckim@khu.ac.kr}
\affiliation{Department of Physics and Research Institute for Basic Sciences, Kyung Hee University, 26 Kyungheedae-ro, Dongdaemun-gu, Seoul 02447, Republic of Korea}
\affiliation{Department of Information Display, Kyung Hee University, 26 Kyungheedae-ro, Dongdaemun-gu, Seoul 02447, Republic of Korea}

%\date{\today}

\begin{abstract}
We report a spin-coating-based approach for forming a percolatively connected graphene oxide (GO) nanosieve on SiO$_2$-patterned sapphire substrates, where the addition of tetrabutylammonium (TBA) to the GO solution significantly improves the uniformity of flake coverage and modulates GaN nucleation behavior. Upon thermal annealing of GO, the resulting reduced graphene oxide (rGO) films exhibit spatially varying coverage, leading to three distinct GaN nucleation outcomes: (i) ELOG-like nucleation on exposed substrate regions, (ii) thru-hole epitaxy (THE)-like nucleation through appropriately thin areas, and (iii) complete nucleation suppression on thickly stacked zones. On spin-coated GO films without TBA, all three behaviors coexist, and undesired ELOG- and no-nucleation modes persist due to uneven coverage. Importantly, these issues cannot be resolved by simply adjusting GO flake concentration, as concentration tuning alone fails to eliminate the formation of locally bare and overly thick regions. In contrast, the addition of TBA results in a more uniform, moderately stacked rGO morphology that suppresses both ELOG- and no-nucleation modes while expanding THE-like nucleation regions. This reshaped nucleation landscape confines GaN growth to areas with engineered percolative transport. The approach offers a scalable, lithography-free route for controlling GaN epitaxy using solution-processable 2D material masks.
\end{abstract}

%\pacs{Valid PACS appear here}

\maketitle

\section{Introduction}
Epitaxial lateral overgrowth (ELOG) is a well-established method for producing high-quality GaN films by enabling selective nucleation within patterned mask openings and promoting lateral overgrowth across masked regions. These growth masks serve to reduce the density of threading dislocations and improve crystal quality, making ELOG essential for optoelectronic and power device applications. However, conventional masks such as SiO$_2$ require complex photolithography or electron-beam lithography, which limits their scalability.

Two-dimensional (2D) materials have recently emerged as promising alternatives to conventional masks due to their atomic thickness, flexibility, and intrinsic ability to support nanoscale openings and transport pathways. Transferred graphene with openings of 100--500~nm has been employed for ELOG.\cite{Xu-APL-111-102105} More recently, our group introduced the concept of \textit{thru-hole epitaxy} (THE), in which precursor diffusion and GaN nucleation occur through nanoscale gaps or openings in transferred 2D multilayers such as h-BN,\cite{Jang-AMI-10-2201406} graphene, and MoS$_2$.\cite{Lee-CGD-22-6995} This mode of growth does not rely on explicitly patterned windows. Monte Carlo simulations further support this mechanism, showing that precursor access through percolative pathways persists in stacked 2D flakes and governs nucleation probability distributions.\cite{Lee-AEM-26-2301654,Lee-CGD-22-6995}

While promising, transferred 2D layers are limited by alignment challenges and scalability issues. In contrast, spin-coating of solution-dispersed flakes offers a scalable, lithography-free alternative. Spin-coated graphene and hexagonal boron nitride (h-BN) nanosheets have been explored as ELOG masks with larger, micron-scale openings.\cite{Zhang-ACSAMI-7-4504} Graphene oxide (GO) is particularly attractive due to its water dispersibility and ability to be thermally reduced into conductive rGO. Spin-coated rGO films have been applied in a variety of contexts, including organic electronics,\cite{Yamada-RSCAdv-9-32940} transparent conductors,\cite{Becerril-ACSNano-2-463} thermal management,\cite{Han-NC-4-1452} flexible devices,\cite{Lu-AMT-3-1700318} and patterning techniques based on spin-coating and transfer.\cite{Guo-ACSNano-4-5749,Reiner-MNL-16-436} Uniform GO coatings have been demonstrated on large-area wafers with high reproducibility,\cite{Yamaguchi-ACSNano-4-524} while thinner GO layers have enabled more pattern-sensitive film morphologies.\cite{Kim-ACSNano-7-8082} However, these studies do not examine the use of spin-coated GO as a selective epitaxy mask, where flake stacking and nanoscale connectivity influence nucleation.

In this study, we demonstrate that modifying the GO solution with tetrabutylammonium (TBA) ions prior to spin-coating leads to the formation of percolatively connected rGO films that are laterally continuous yet permeable to precursor transport through nanoscale gaps. These TBA-modified films effectively guide GaN nucleation via transport-limited pathways, enabling thru-hole epitaxy (THE)-like behavior without requiring lithographically defined openings. Through systematic investigation, we found that when unmodified GO is used as a mask, achieving suitable coverage for selective-area epitaxy is challenging: lower flake concentrations leave large areas of the substrate exposed, while higher concentrations result in flake aggregation and multi-stacked regions that inhibit nucleation. Critically, even at these higher concentrations, uncovered regions persist due to non-uniform flake dispersion, leading to a coexistence of three distinct nucleation behaviors—ELOG-like growth on exposed areas, nucleation suppression in overstacked regions, and sporadic THE-like growth in intermediate zones. These issues cannot be resolved by adjusting flake concentration alone, underscoring the need for additive-assisted processing to achieve uniform, moderately stacked rGO coverage suitable for controlled nucleation. It is well known that obtaining uniform GO films by spin-coating without additives is very challenging,\cite{Yamaguchi-ACSNano-4-524} but TBA has been previously shown to improve GO dispersion and suppress interflake interactions, producing more uniform or mosaic-like monolayer films.\cite{Kim-ACSNano-7-8082} However, its application in selective-area epitaxy—particularly as a solution-processed alternative to conventional THE masks—has remained underexplored. Our approach provides a scalable, lithography-free route to spatially modulated GaN nucleation by tuning GO flake stacking and coverage, thereby expanding the utility of 2D materials in epitaxy and introducing a new paradigm for self-organized mask formation.

\section{Experimental methods}

To evaluate the influence of GO film uniformity on GaN nucleation behavior, we employed a SiO$_2$-patterned sapphire substrate rather than a bare substrate. This patterned design, consisting of a hexagonal array of circular openings that expose the underlying sapphire, was critical for enabling quantitative and statistical analysis of GaN domain distribution. Each opening serves as an independent unit, allowing consistent and repeatable assessment of nucleation events. This spatial regularity not only reduces the randomness inherent to bare substrates but also permits binary classification of nucleation outcomes (e.g., presence or absence of a GaN domain). By analyzing large numbers of equivalent openings, we systematically compared nucleation behavior across samples and identified trends associated with TBA-assisted rGO morphology. Without such patterning, the uncontrolled nature of nucleation on bare surfaces would obscure these effects and render meaningful statistical comparisons impractical. 

A 50-nm-thick SiO$_2$ layer was deposited onto a $c$-plane sapphire wafer, and a hexagonal close-packed array of circular openings, each 4~$\mu$m in diameter, was defined via photolithography. This pattern exposed bare sapphire within the openings, while the surrounding regions remained SiO$_2$-coated. Spin-coated rGO flakes covered both the exposed sapphire and the SiO$_2$ regions. During subsequent GaN growth, nucleation occurred on both surfaces; however, only the nuclei formed on the rGO-covered sapphire regions were crystallographically aligned with the substrate (as confirmed by the alignment of well-developed side facets), initiating epitaxial growth. In contrast, the nuclei observed on the SiO$_2$ regions were not epitaxially aligned. Therefore, our analysis of nucleation behavior focuses on domains formed on the sapphire substrate, where epitaxial alignment governs the resulting GaN microstructure.

GO flake suspensions were prepared at two concentrations—0.3~mg/mL and 1.0~mg/mL—in deionized (DI) water. For TBA-modified GO (TGO), 0.02~mL of a 1~wt\% aqueous tetrabutylammonium hydroxide (TBAOH) solution was added to 8~mL of a 4~mg/mL GO solution. The mixture was then diluted with DI water to yield a final GO concentration of 0.3~mg/mL. The resulting TBAOH concentration was 1.9~$\mu$g/mL—negligible relative to GO—and the solution is thus referred to as TGO 0.3~mg/mL, based on the GO concentration. All GO and TGO solutions were spin-coated onto the prepared sapphire substrates, followed by thermal annealing at 900$^\circ$C under a nitrogen atmosphere to convert GO into reduced graphene oxide (rGO).

Subsequently, GaN was grown directly on the rGO-coated substrates using hydride vapor phase epitaxy (HVPE) at 1020$^\circ$C for one minute, without the use of a low-temperature GaN buffer layer. Identical growth conditions were applied to all rGO samples to ensure a consistent and fair comparison of nucleation and growth behaviors across the different morphologies.

\section{Results and discussion}
\begin{figure}
\includegraphics[width=1.0\columnwidth]{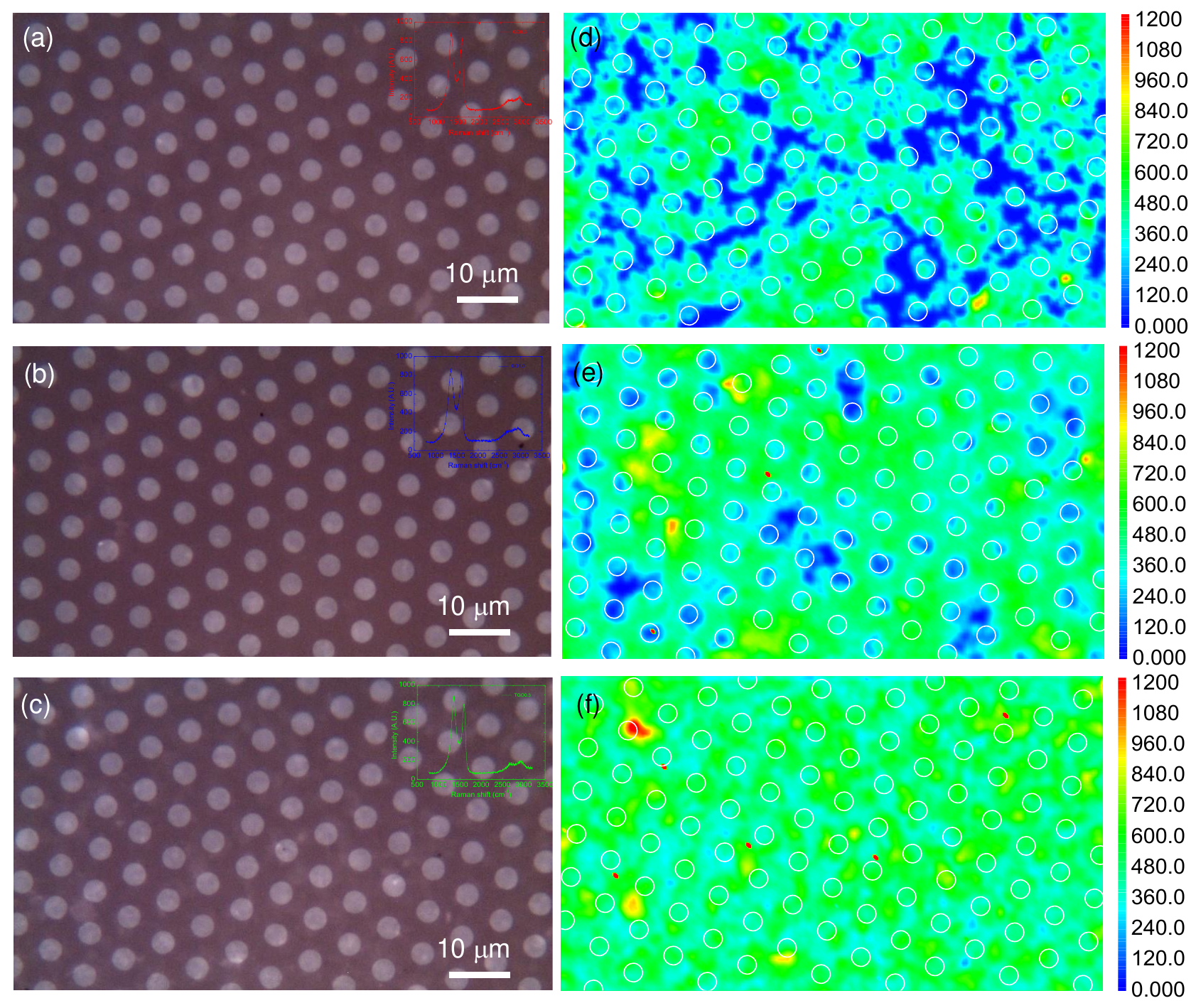}
\caption{Optical microscopy images and corresponding Raman G-peak intensity maps of SiO$_2$-patterned $c$-plane sapphire substrates spin-coated with graphene oxide (GO) solutions under varying concentrations and TBA conditions: (a, d) 0.3~mg/mL GO without TBA (Sample I), (b, e) 1.0~mg/mL GO without TBA (Sample II), and (c, f) 0.3~mg/mL GO with TBA (Sample III). All samples were thermally annealed at 900$^\circ$C in a nitrogen atmosphere to reduce GO into rGO. Insets in (a–c) show representative Raman spectra collected from within the circular sapphire openings, exhibiting the characteristic rGO peaks: D ($\sim$1350~cm$^{-1}$), G ($\sim$1600~cm$^{-1}$), and 2D ($\sim$2700~cm$^{-1}$). G-peak intensity maps in (d–f) visualize local variations in rGO coverage and thickness across the patterned surfaces, with particular emphasis on inter-opening variation.}
\label{OM-Raman-As-is-rGO}
\end{figure}

To evaluate the feasibility of using rGO films as effective masks for thru-hole epitaxy, it is important to examine their spatial uniformity and their potential to support percolative transport. The film must (1) continuously cover the substrate to prevent conventional ELOG from occurring via explicitly exposed regions, and (2) remain sufficiently thin or structurally percolative to allow precursor species to access the underlying substrate through nanoscale gaps. Due to the inherent characteristics of spin-coating, thickness variation is inevitable across a sample. In this study, Raman mapping was used as a proxy to assess spatial variation in rGO film thickness, leveraging the known correlation between Raman G-peak intensity and local carbon content. Although the G-peak intensity does not provide a direct thickness measurement, it serves as a qualitative indicator for evaluating average flake density and coverage across the patterned circular openings. 

SiO$_2$-patterned $c$-plane sapphire substrates were spin-coated with GO solutions of varying concentration and TBA content, then thermally annealed to produce rGO films. The resulting samples are labeled as Sample I (0.3~mg/mL GO without TBA), Sample II (1.0~mg/mL GO without TBA), and Sample III (0.3~mg/mL GO with TBA). Optical microscopy images [Fig.~\ref{OM-Raman-As-is-rGO}(a–c)] reveal limited contrast variation among the three samples. However, representative Raman spectra collected from within the circular sapphire openings [insets of Fig.~\ref{OM-Raman-As-is-rGO}(a–c)] exhibit the characteristic D ($\sim$1350~cm$^{-1}$), G ($\sim$1600~cm$^{-1}$), and 2D ($\sim$2700~cm$^{-1}$) bands, indicating that thermal treatment produces carbon-rich films with spectral features characteristic of rGO.

The Raman G-peak intensity maps [Fig.~\ref{OM-Raman-As-is-rGO}(d–f)] reveal marked differences in rGO film distribution. In Sample I, low G-peak intensity (blue regions) dominates within the openings, suggesting sparse flake coverage and a high probability of exposed substrate. In Sample II, increased overall intensity implies greater coverage, although some uncovered zones still persist. In contrast, Sample III shows uniformly high G-peak intensity across the patterned area, with the absence of blue regions indicating continuous coverage. There is, however, still morphologically non-uniform stacking. These observations suggest that TBA promotes a more homogeneous rGO film morphology, minimizing both uncovered and overstacked regions. While Raman mapping alone does not confirm the formation of percolative gaps, the successful nucleation and growth of GaN through these films—as shown in a later section—despite their stacked nature, supports the existence of percolative nanoscale transport pathways. This interpretation is consistent with our prior experimental and computational studies,\cite{Jang-AMI-10-2201406,Lee-CGD-22-6995,Lee-AEM-26-2301654} which demonstrate the persistence of percolative networks in stacked 2D materials.

\begin{figure}
\includegraphics[width=1.0\columnwidth]{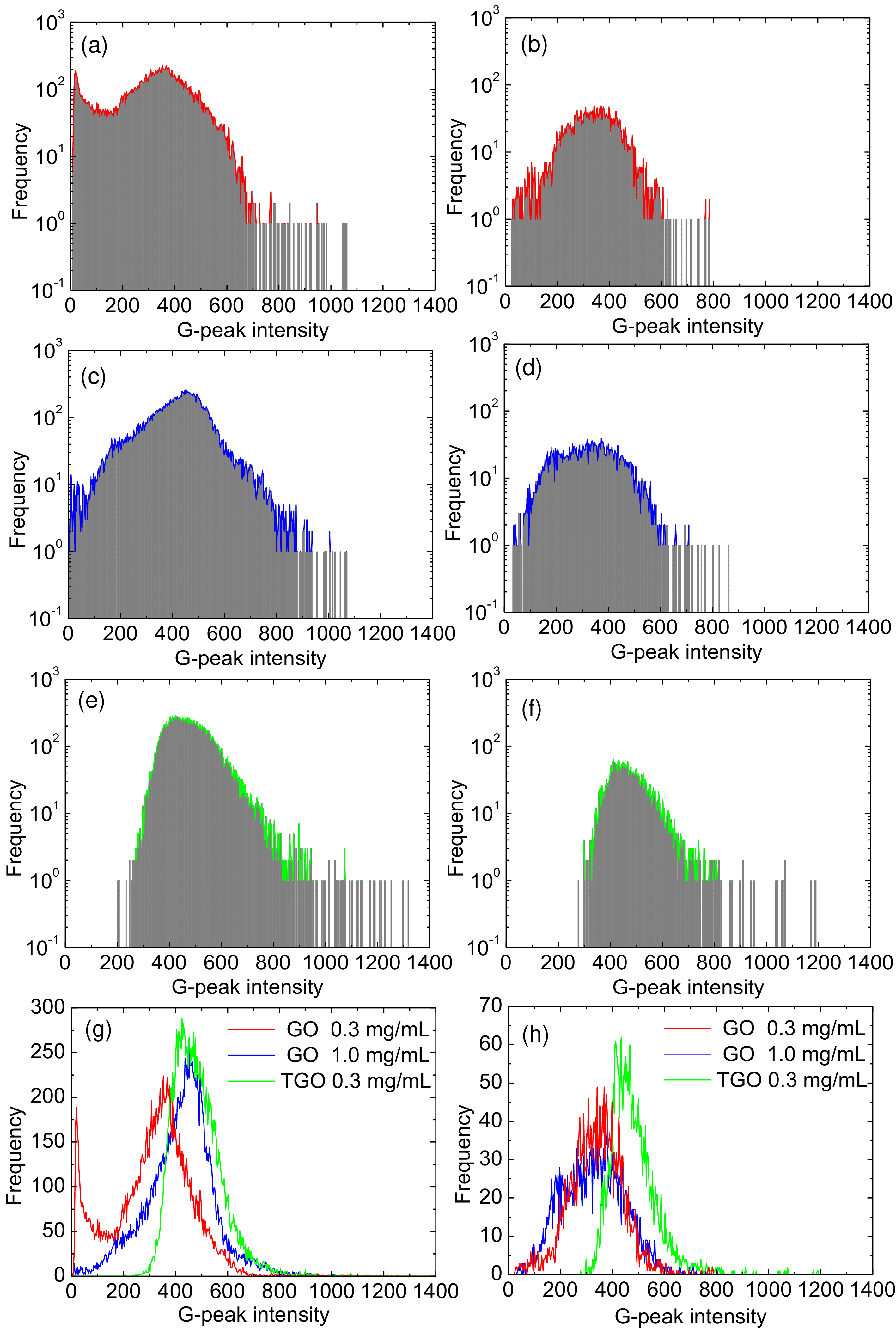}
\caption{Histograms of Raman G-peak intensity distributions for Samples I–III. (a, c, e) G-peak intensity distributions over the entire measured area for Sample I (0.3~mg/mL GO without TBA), Sample II (1.0~mg/mL GO without TBA), and Sample III (0.3~mg/mL GO with TBA), respectively. (b, d, f) Corresponding histograms restricted to the circular sapphire opening regions. (g, h) Comparative plots of G-peak intensity distributions for all three samples over (g) the full area and (h) the circular openings, plotted on a linear scale. These comparisons highlight variations in rGO film coverage and spatial uniformity as a function of GO concentration and TBA treatment.}
\label{Histogram-before-GaN-growth}
\end{figure}

The influence of GO flake concentration and TBA addition on rGO film uniformity becomes more evident through statistical analysis of the Raman G-peak intensity distributions, as shown in Fig.~\ref{Histogram-before-GaN-growth}. Alongside the individual histograms in Fig.~\ref{Histogram-before-GaN-growth}(a), (c), and (e), the overlaid comparison in Fig.~\ref{Histogram-before-GaN-growth}(g) highlights two key observations: (1) Sample II (1.0~mg/mL GO without TBA) exhibits a peak at higher G-peak intensity compared to Sample I (0.3~mg/mL GO without TBA), consistent with increased carbon coverage, but retains a long low-intensity tail indicative of sparse regions; and (2) Sample III (0.3~mg/mL GO with TBA) shows a similar peak position to Sample II but lacks the low-intensity tail entirely. These trends suggest that increasing flake concentration alone does not eliminate local coverage deficiencies, whereas the addition of TBA effectively suppresses these sparse zones. Based on the earlier discussion, an ideal G-peak intensity distribution for a uniform and functional rGO mask should exhibit a narrow profile without a low-intensity tail. These results demonstrate that TBA modification is more effective than simply increasing GO concentration in achieving uniform rGO coverage.

Since GaN epitaxy is expected to occur only within the circular sapphire openings defined by the SiO$_2$ mask—provided that the rGO film permits precursor access—we examined the G-peak intensity distributions restricted to these regions [Fig.\ref{Histogram-before-GaN-growth}(b), (d), (f), (h)]. Comparing the full-area histograms [Fig.\ref{Histogram-before-GaN-growth}(a), (c), (e), (g)] to those within the openings reveals that Sample III uniquely exhibits scaled-down but similarly shaped distributions in both cases, indicating that the rGO coverage is homogeneously maintained even at the local scale. In contrast, Samples I and II show noticeable divergence between the overall and opening-restricted distributions, reflecting spatial non-uniformity within the functional epitaxial regions. These results support the view that TBA promotes uniform and continuous rGO film morphology across both global and local scales, which is critical for enabling localized precursor transport and, ultimately, successful thru-hole epitaxy.

\begin{figure}
\includegraphics[width=1.0\columnwidth]{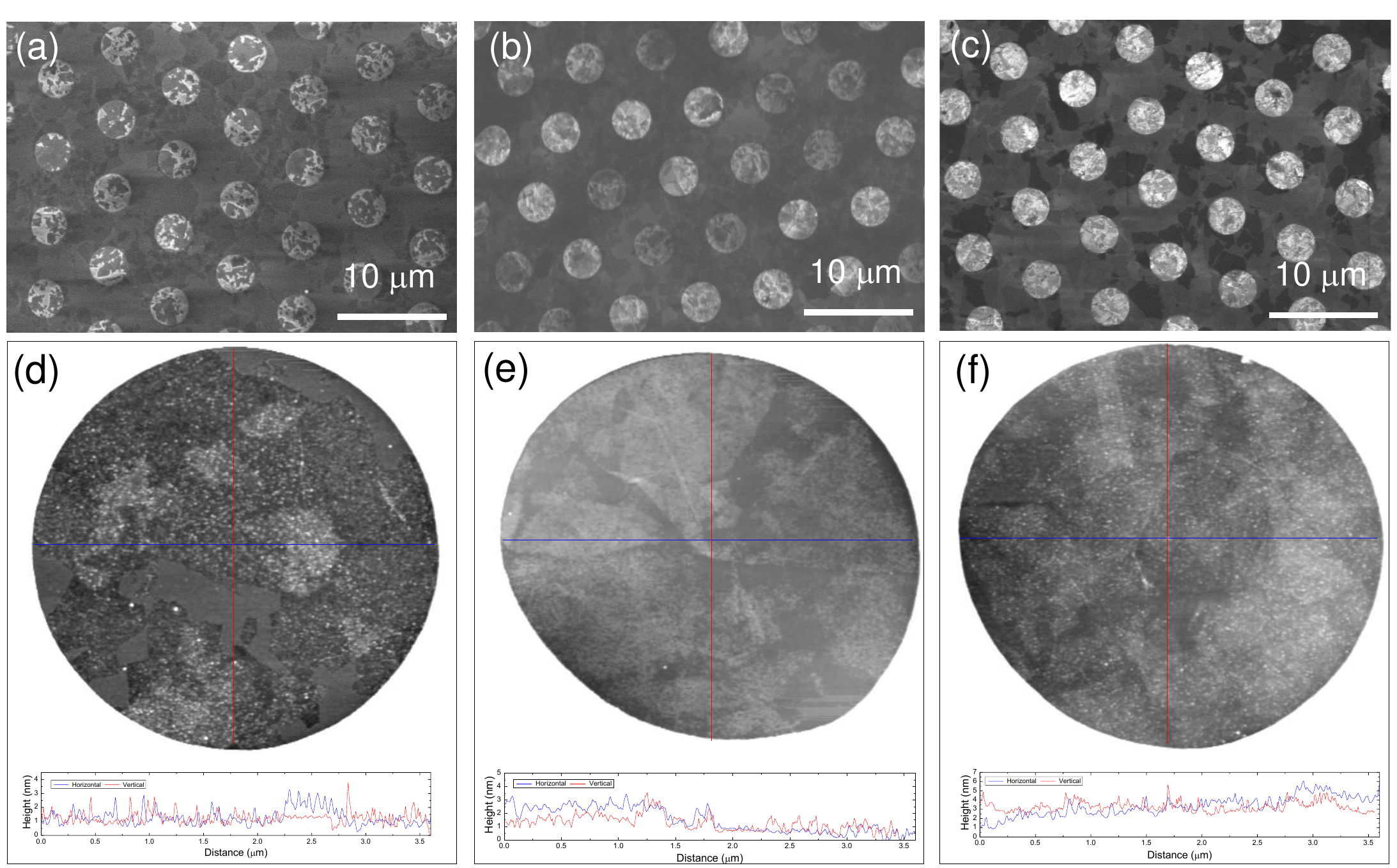}
\caption{SEM and AFM topography images of SiO$_2$-patterned $c$-plane sapphire substrates spin-coated with GO flakes and thermally annealed: (a, d) 0.3~mg/mL GO without TBA (Sample I), (b, e) 1.0~mg/mL GO without TBA (Sample II), and (c, f) 0.3~mg/mL GO with TBA (Sample III). The AFM images were acquired from individual circular openings. Height profiles corresponding to the blue and red lines in the topography panels are shown below each image, illustrating intra-opening variations in rGO flake stacking.}
\label{SEM-AFM-As-is-rGO}
\end{figure}

To investigate the distribution of stacked rGO flakes within individual circular openings in greater detail, we performed SEM imaging and AFM topography measurements. The SEM images in Fig.~\ref{SEM-AFM-As-is-rGO}(a–c) show that rGO flakes are present both inside and outside the circular openings defined by the SiO$_2$ pattern. Notably, contrast variation within the openings suggests significant intra-opening variation in total flake thickness. While the SEM images qualitatively confirm the presence of rGO and its morphological heterogeneity, further insight into the degree of thickness variation is provided by AFM analysis.

AFM topography images of selected openings for each sample [Fig.\ref{SEM-AFM-As-is-rGO}(d–f)] reveal clear differences in intra-opening thickness distribution. In Samples I and II [Fig.\ref{SEM-AFM-As-is-rGO}(d), (e)], relatively flat and low-roughness regions are observed. Although the corresponding height profiles indicate that these areas are not completely bare, they are covered only by ultra-thin rGO layers—potentially thin enough to locally expose or insufficiently block the underlying substrate. Such weakly masked or partially uncovered regions increase the likelihood of direct nucleation on the substrate, resembling epitaxial lateral overgrowth (ELOG) behavior. In contrast, Sample III [Fig.~\ref{SEM-AFM-As-is-rGO}(f)] exhibits a more uniformly textured topography with no extended flat segments, consistent with more homogeneous and complete intra-opening flake stacking. This suggests that TBA promotes better coverage within the openings, thereby minimizing the chances for ELOG-like nucleation. Instead, the improved morphology in Sample III is expected to support more consistent percolative pathways, enhancing the reliability of thru-hole epitaxy during GaN growth.

\begin{figure}
\includegraphics[width=1.0\columnwidth]{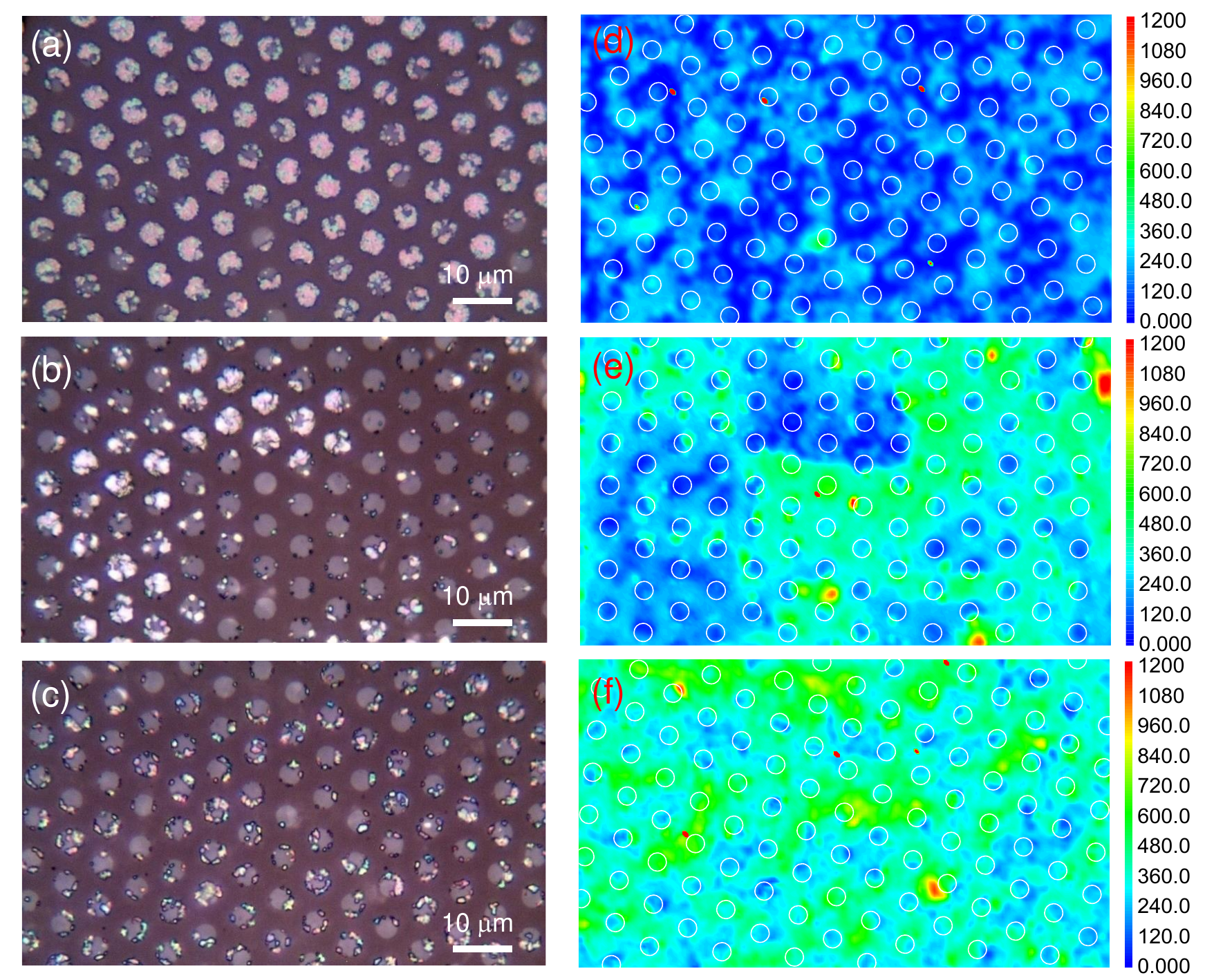}
\caption{Optical microscopy images and Raman G-peak intensity maps of GaN domains grown on SiO$_2$-patterned $c$-plane sapphire substrates spin-coated with GO flakes and thermally annealed under three different conditions: (a, d) 0.3~mg/mL GO without TBA (Sample IV), (b, e) 1.0~mg/mL GO without TBA (Sample V), and (c, f) 0.3~mg/mL GO with TBA (Sample VI). Panels (a–c) show the distribution of GaN domains after growth, while panels (d–f) display corresponding Raman G-peak intensity maps. The G-peak maps highlight the extent of rGO degradation and its spatial correlation with the GaN domain coverage.}
\label{Raman-map-after-GaN-growth}
\end{figure}

To investigate how the thickness and uniformity of stacked rGO flakes influence GaN domain formation, we performed GaN growth for one minute on SiO$_2$-patterned $c$-plane sapphire substrates coated with GO at varying concentrations, with and without TBA modification. All samples were thermally annealed to reduce the GO prior to GaN growth. The resulting samples are referred to as Sample IV (0.3~mg/mL GO without TBA), Sample V (1.0~mg/mL GO without TBA), and Sample VI (0.3~mg/mL GO with TBA). Figure~\ref{Raman-map-after-GaN-growth} shows optical images of the GaN domains formed within the circular openings, alongside Raman G-peak intensity maps acquired after growth.

For Sample IV [Fig.\ref{Raman-map-after-GaN-growth}(a)], GaN domains are observed in nearly all circular openings, suggesting effective precursor access to the substrate due to the low GO concentration. In contrast, Sample V [Fig.\ref{Raman-map-after-GaN-growth}(b)] exhibits significantly reduced and spatially non-uniform domain coverage—some openings contain dense GaN domains, while others are completely devoid of growth. This behavior indicates that simply increasing GO concentration leads to excessive flake stacking, which impedes consistent precursor diffusion. Sample VI [Fig.~\ref{Raman-map-after-GaN-growth}(c)], prepared with TBA, shows more uniform GaN distribution across openings, although the overall coverage is slightly lower compared to Sample IV.  These trends are consistent with the G-peak intensity distributions shown in Fig.~\ref{Raman-map-after-GaN-growth}(d),(e),(f). Regions with higher G-peak intensity—indicative of thicker rGO coverage—correlate with suppressed GaN nucleation, reinforcing the interpretation that overstacked flakes hinder precursor access. 

\begin{figure}
\includegraphics[width=1.0\columnwidth]{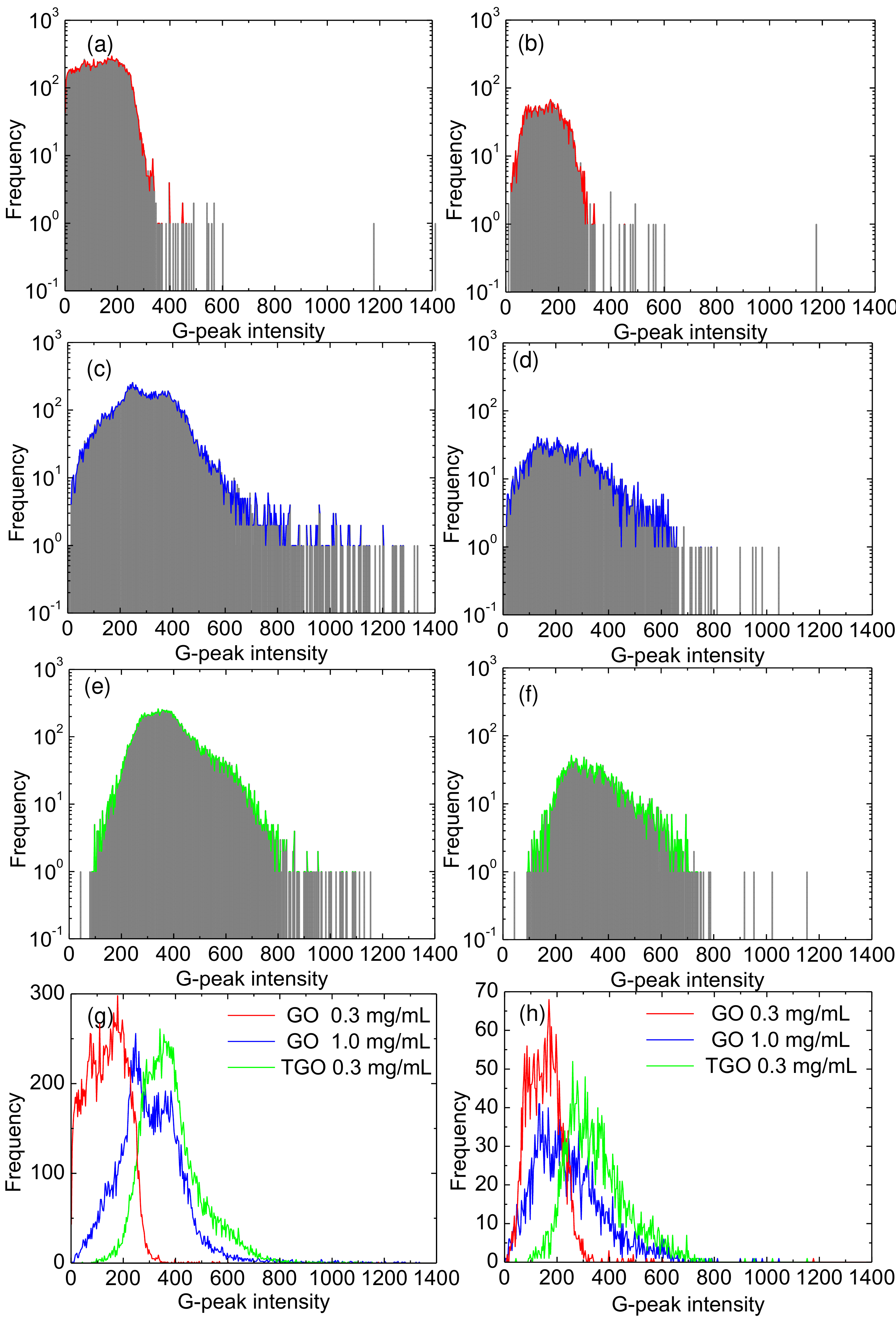}
\caption{Histograms of Raman G-peak intensity distributions after GaN growth for Samples IV–VI. (a, c, e) G-peak intensity distributions over the entire scanned area for Sample IV (0.3~mg/mL GO without TBA), Sample V (1.0~mg/mL GO without TBA), and Sample VI (0.3~mg/mL GO with TBA), respectively. (b, d, f) Corresponding histograms restricted to the circular sapphire openings. Comparative plots of all three samples are shown on a linear scale for (g) the full area and (h) the circular openings, highlighting changes in rGO film coverage and integrity following GaN growth.}
\label{Histogram-after-GaN-growth}
\end{figure}

Further differences in rGO behavior after GaN growth are evident from the G-peak intensity distributions shown in Fig.~\ref{Histogram-after-GaN-growth}. One of the most prominent changes across all samples is a marked shift toward lower values, indicating the derease in overall G-peak intensity due to the degradation or thinning of the rGO film during GaN growth. This trend is especially pronounced in Sample IV (0.3~mg/mL GO without TBA), where the initial rGO coverage was thinner and more spatially non-uniform, rendering it more susceptible to thermal decomposition during growth.  Previous studies have shown that the thermal stability of transferred graphene strongly depends on the substrate: graphene on sapphire remains relatively stable, whereas graphene on GaN templates decomposes due to interfacial reactions.\cite{Park-AMI-6-1900821} In our study, rGO films—even on sapphire—undergo significant degradation under atmospheric-pressure growth conditions, likely due to differences in the thermal environment compared to prior studies conducted under reduced pressure (100–200 Torr).\cite{Park-AMI-6-1900821} Additionally, the intrinsically lower structural integrity of rGO—characterized by a high density of oxygen-containing defects, lower crystallinity, and a large edge-to-basal area ratio—may further reduce its thermal stability and promote irreversible decomposition at elevated temperatures.\cite{Eigler-ACIE-53-7720}  

These results reinforce earlier findings that both the average thickness and spatial uniformity of the rGO flakes—particularly within the circular openings—are highly sensitive to GO flake concentration and the presence of TBA. Intra-opening variations, as revealed by AFM topography in Fig.~\ref{SEM-AFM-As-is-rGO}, further suggest that such local thickness fluctuations may determine whether GaN growth proceeds via direct access to exposed substrate regions or through confined percolative transport pathways.

\begin{figure*}
\includegraphics[width=2.0\columnwidth]{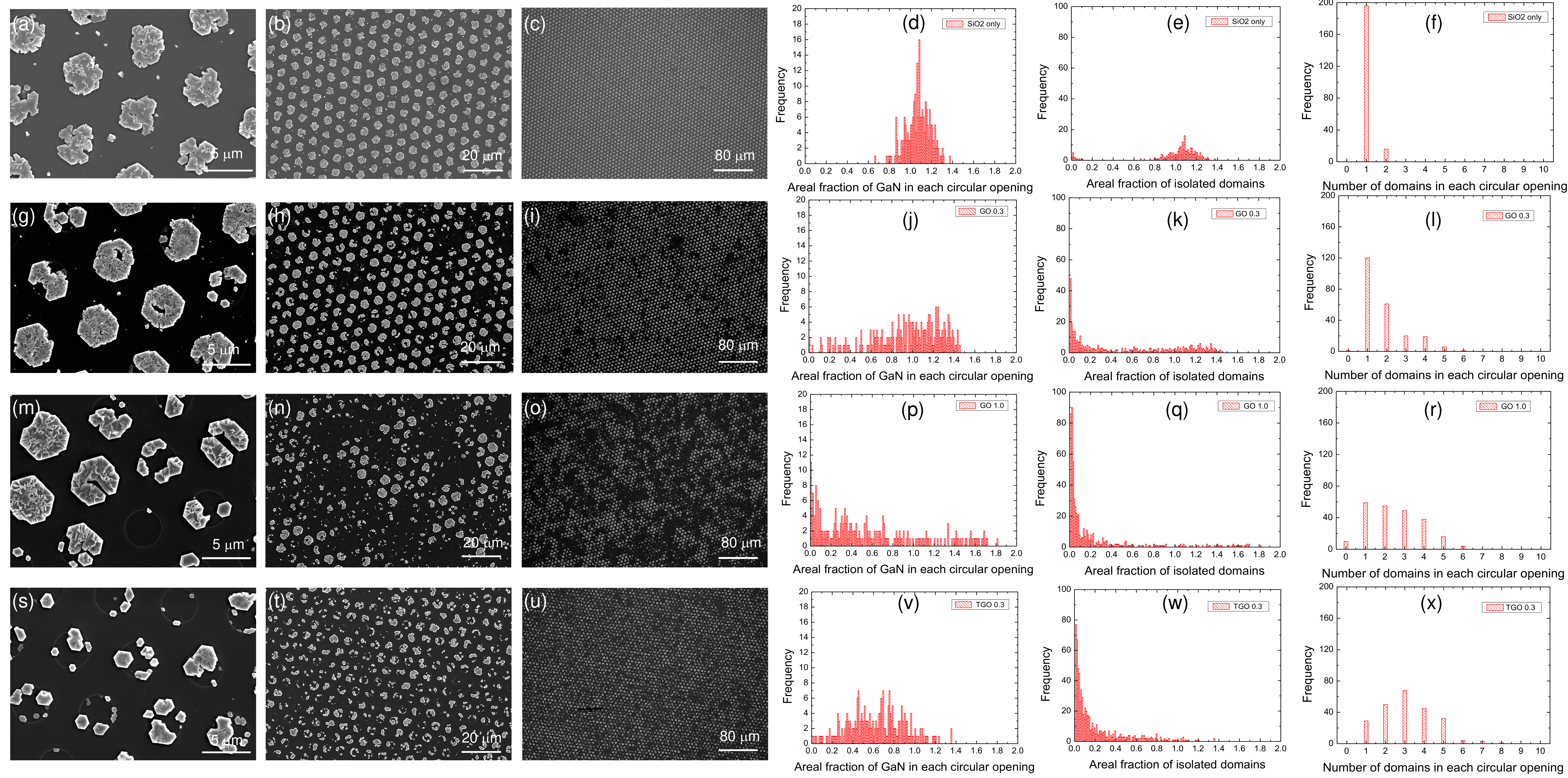}
\caption{SEM images of GaN domains grown for one minute on SiO$_2$-patterned $c$-plane sapphire under four different surface conditions, shown at magnifications of ×5,000, ×1,000, and ×100: (a–c) uncoated SiO$_2$-patterned substrate, (g–i) 0.3~mg/mL GO without TBA, (m–o) 1.0~mg/mL GO without TBA, and (s–u) 0.3~mg/mL GO with TBA.  (d–f), (j–l), (p–r), and (v–x) show histograms of the areal fraction of GaN in each circular opening, the areal fraction of isolated domains, and the number of domains per opening, respectively, for each case. The areal fraction is defined as the ratio of the GaN-covered area to the area of a circular opening; that is, an areal fraction of one indicates that the GaN area is equal to the area of the circular opening. Notably, for the uncoated SiO$_2$-patterned substrate, single merged domains were observed in almost every circular opening, consistent with uniform ELOG behavior in the absence of a masking layer.}
\label{GaN-domain-distribution}
\end{figure*}

To further distinguish between epitaxial lateral overgrowth (ELOG) and thru-hole epitaxy (THE), we conducted a statistical analysis of GaN domain size and areal coverage for samples grown on both rGO-uncoated and rGO-coated substrates under different processing conditions, as shown in Fig.~\ref{GaN-domain-distribution}. SEM images at multiple magnifications reveal that GaN domains are densely distributed in most circular openings across all cases, but domain size, coalescence behavior, and spatial uniformity vary markedly with rGO morphology.

For the rGO-uncoated substrate, the areal fraction of domains [Fig.~\ref{GaN-domain-distribution}(d)], the areal fraction of isolated domains [Fig.~\ref{GaN-domain-distribution}(e)], and the number of domains per opening [Fig.~\ref{GaN-domain-distribution}(f)] all exhibit well-defined peak distributions. This indicates that each circular opening typically hosts a single, uniformly sized merged domain. Because the entire opening area is accessible for nucleation, a conventional ELOG-like nucleation process occurs uniformly within each opening. However, samples prepared with unmodified GO (Samples IV and V) exhibit characteristics of both ELOG and THE. In Sample IV (0.3~mg/mL GO), some openings are fully occupied by large, merged domains—indicative of ELOG nucleation on exposed substrate regions—while others remain completely empty, suggesting THE-like suppression due to excessive flake stacking that inhibits precursor diffusion [Figs.~\ref{GaN-domain-distribution}(j)--(l)]. Sample V (1.0~mg/mL GO) shows a similar trend with slightly improved coverage, but still exhibits heterogeneous nucleation behavior consistent with spatially nonuniform rGO films containing both undercovered and overcovered regions [Figs.~\ref{GaN-domain-distribution}(p)--(r)].  In contrast, Sample VI (0.3~mg/mL GO with TBA) displays a more uniform distribution of smaller, unmerged domains [Figs.~\ref{GaN-domain-distribution}(v)--(x)]. These features suggest a substantial kinetic delay in both nucleation and lateral growth, which cannot be attributed to external conditions, as all samples were grown under identical HVPE parameters. Instead, this delay indicates more homogeneous rGO coverage, with precursor transport confined to percolative pathways. Although THE in this context does not result in deterministic nucleation positions, the suppression of domain coalescence and the absence of either fully unoccupied or overmerged openings strongly support a thru-hole epitaxy mechanism enabled by a percolatively connected rGO nanosieve. Supporting this, Raman G-peak maps confirm the presence of rGO-free and overstacked regions in the GO-only samples, while such features are absent in the TBA-modified case.

In addition to differences in domain size and merging behavior, we observed distinctions in the extent of side-facet formation between samples. In particular, Sample VI, where thru-hole epitaxy (THE) is more dominant, exhibits smaller but sharply faceted domains. This observation is consistent with kinetic Wulff theory, which predicts that smaller domains evolve more quickly toward their equilibrium or kinetically preferred shapes, resulting in well-defined side facets within short growth durations.\cite{Jue-APL-104-092110, Lee-APL-104-182105}  We also note that the Sample IV and V—both exhibiting mixed nucleation behavior—show more pronounced faceting than the SiO$_2$-patterned only sample. This is also likely due to a lower nucleation site density in the GO-coated samples, which provides more lateral space and time for individual domains to develop prior to merging. In contrast, the higher density of nucleation events in the SiO$_2$-patterned sample promotes rapid domain coalescence, limiting facet development in such a short growth duration of one minute.

This behavior suggests that TBA modification leads to a more continuous but permeable rGO film, enabling consistent precursor diffusion through nanoscale pathways across all openings. These observations support the interpretation that TGO promotes a more uniform thru-hole epitaxy process, while the GO-only samples exhibit a mixed nucleation landscape dominated by local variations in rGO coverage.  This approach may be extended to other epitaxial systems—such as AlN, ZnO, or transition metal dichalcogenides—where selective nucleation, defect reduction, or nanoscale confinement of precursor transport is desired.  It is important to note that, in this work, thru-hole epitaxy does not result in sharply localized or periodic nucleation sites, as the percolative pathways are randomly distributed and not lithographically defined.

\section{Conclusion}
We demonstrated that reduced graphene oxide (rGO) films, formed by spin-coating graphene oxide (GO) flakes onto SiO$_2$-patterned sapphire substrates and thermally reducing them, can serve as effective self-organized masks for GaN epitaxy. The incorporation of tetrabutylammonium (TBA) ions significantly improved the uniformity of rGO flake stacking within the circular openings. A key finding is that the local stacking configuration and continuity of the rGO film directly influence GaN nucleation behavior: discontinuous regions lead to conventional epitaxial lateral overgrowth (ELOG), while more continuous and uniform films promote thru-hole epitaxy (THE) via nanoscale percolative pathways. This is supported by (1) successful GaN nucleation through multilayered rGO and (2) delayed nucleation and lateral growth kinetics in TBA-treated samples. Raman, SEM, and AFM analyses consistently corroborate this correlation. These results establish TBA-assisted spin-coating as a simple, scalable method to tailor rGO morphology and guide GaN growth without lithography, offering practical relevance for low-cost, selective epitaxy in optoelectronic and power devices.

While TBA has previously been used to improve GO dispersion for uniform or mosaic-like films,\cite{Yamaguchi-ACSNano-4-524,Kim-ACSNano-7-8082} this study uniquely demonstrates its impact on growth-selective epitaxy. By tuning flake stacking morphology, we enable a solution-processable route to percolatively connected nanosieve masks that regulate nucleation via nanoscale precursor transport—an application not explored in earlier work.

Looking ahead, further exploration of flake size distribution, degree of reduction, and stacking parameters may provide deeper insight into precursor-accessible pathways and enhance nucleation control. This approach opens new directions for scalable rGO-based nanosieve masks in advanced epitaxial systems.

\section{acknowledgement}
This work was supported by the National Research Foundation of Korea(NRF) grant funded by the Korea government (MSIT) (RS-2021-NR060087, RS-2023-00240724) and through Korea Basic Science Institute (National research Facilities and Equipment Center) grant (2021R1A6C101A437) funded by the Ministry of Education.

%\nocite{*}
%\bibliography{reference}% Produces the bibliography via BibTeX.


\end{document}